\documentclass{aa}  
\usepackage{graphicx}
\usepackage{multirow}
\usepackage{subcaption}
\usepackage[normalem]{ulem}

\usepackage{color}

\definecolor{tom}{rgb}{0.0, 0., 1.0}
\definecolor{vaibhav}{rgb}{0.1, 0.5, 0.2}

\definecolor{lianne}{rgb}{0.7, 0.0, 0.5}

\definecolor{darkgreen}{RGB}{0,153,0}

\makeatletter
\newcommand*{\rom}[1]{\expandafter\@slowromancap\romannumeral #1@}
\makeatother
\newcommand{\boldnabla}{\mbox{\boldmath$\nabla$}}
\usepackage{siunitx}
\newcommand{\Alfvenic}{Alfv\'enic }
\newcommand{\Alfven}{Alfv\'en }
\usepackage[utf8]{inputenc}
\usepackage{fourier} 
\usepackage{array}
\usepackage{makecell}

\usepackage{placeins}

\begin{document}

\title{Investigating coronal wave energy estimates using synthetic non-thermal line widths}
\author{L.~E. Fyfe \inst{1} , T.~A. Howson \inst{1}, I. De Moortel \inst{1,2}, V. Pant \inst{3} and T. Van Doorsselaere \inst{4}}

\authorrunning{Fyfe et al.}

\institute{School of Mathematics and Statistics, University of St. Andrews, St. Andrews, Fife, KY16 9SS, U.K. \and Rosseland Centre for Solar Physics, University of Oslo, PO Box 1029  Blindern, NO-0315 Oslo, Norway \and Aryabhatta Research Institute of Observational Sciences, Nainital, Uttarakhand
263001, India \and Centre for mathematical Plasma Astrophysics, Department of Mathematics, KU Leuven, Celestijnenlaan 200B, Leuven, Belgium}

\abstract{}
{Estimates of coronal wave energy remain uncertain as a large fraction of the energy is likely hidden in the non-thermal line widths of emission lines. In order to estimate these wave energies, many previous studies have considered the root mean squared wave amplitudes to be a factor of $\sqrt{2}$ greater than the non-thermal line widths.  However, other studies have used different factors. To investigate this problem, we consider the relation between wave amplitudes and the non-thermal line widths within a variety of 3D magnetohydrodynamic (MHD) simulations.}
{We consider the following 3D numerical models: \Alfven waves in a uniform magnetic field, transverse waves in a complex braided magnetic field, and two simulations of coronal heating in an arcade. We applied the forward modelling code FoMo to generate the synthetic emission data required to analyse the non-thermal line widths.}
{Determining a single value for the ratio between the non-thermal line widths and the root mean squared wave amplitudes is not possible across multiple simulations. It was found to depend on a variety of factors, including line-of-sight angles, velocity magnitudes, wave interference, and exposure time. Indeed, some of our models achieved the values claimed in recent articles while other more complex models deviated from these ratios.}
{To estimate wave energies, an appropriate relation between the non-thermal line widths and root mean squared wave amplitudes is required. However, evaluating this ratio to be a singular value, or even providing a lower or upper bound on it, is not realistically possible given its sensitivity to various MHD models and factors. As the ratio between wave amplitudes and non-thermal line widths is not constant across our models, we suggest that this widely used method for estimating wave energy is not robust.}




\maketitle


\section{Introduction}\label{sec:introduction}



It is well known that the solar corona is heated up to millions of degrees. The primary mechanisms proposed to achieve this heating can be separated into two classes: the dissipation of stored magnetic energy and the dissipation of magnetohydrodynamic (MHD) waves \citep[see, for example,][for reviews on coronal heating theories]{ParnellDeMoortel2012,Arregui2015,DeMoortelBrowning2015,Klimchuk2015,VanDoorsselaereNakariakov2020}. In recent years, due to higher spatio-temporal resolution of imaging and spectroscopic instruments, MHD waves have been shown to be ubiquitous within the solar atmosphere. One signature of these waves is the non-thermal broadening of emission lines \citep[e.g.][]{Hollweg1973,VanDoorsselaereNakariakov2008}. Using the slit spectrograph aboard Skylab, the broadening of transition region emission lines, as well as the broadening of the spectra in quiet Sun regions and coronal holes have been observed \citep[e.g.][]{DoscheckFeldmanBohlin1976,DoscheckFeldmanVanhoosier1976,FeldmanDoschek1976}. Subsequently, \citet{HasslerRottman1990} detected the broadening of the transition region and coronal emission lines and concluded that the most likely cause was waves in the corona. Some other studies found that non-thermal broadening varies with height through the solar atmosphere. For example, \cite{DoyleBanerjee1998} found an increase in the Si \rom{8} non-thermal line width with increasing altitude above the solar limb, whereas \cite{HahnLandi2012} reported a decrease in line width at relatively low heights in coronal holes.

Counter-propagating waves are thought to be present in the solar atmosphere and can cause turbulence. Such turbulence can go on to broaden emission lines \citep[e.g.][]{TomczykMcIntosh2009,LiuMcIntosh2014,MortonTomczyk2015, VanBallegooijenAsgariTarghi2017}. Although, the non-thermal broadening of emission lines is not necessarily due to the unresolved temporal Doppler velocity amplitudes caused by MHD waves, other solar phenomena can influence the non-thermal line widths as well. These include plasma upflows and plumes near magnetic footpoints \citep[e.g.][]{DePontieuMcIntosh2010,TianMcIntoshDePontieu2011,TianMcIntosh2012} and larger scale upflows within coronal holes \citep[e.g.][]{McIntoshLeamon2011, TianMcIntoshHabbal2011}.

Enhanced non-thermal line widths are a signature of multiple unresolved plasma flows along the line-of-sight (LOS). Hence, they can account for the discrepancy between the true wave energy and the observed wave energy attained from Doppler velocities \citep[e.g.][]{McIntoshDePontieu2012,PantMagyar2019}. In a previous study, \citet{DeMoortelPascoe2012} present a 3D model of transverse waves propagating along multiple loop strands. These waves were generated by a lower boundary driver designed to mimic random footpoint motions. The authors found that by estimating the kinetic energy using the LOS Doppler velocities, it fails to capture at least 60\% of the total kinetic energy in the simulation and hence it is essential to include the enhanced non-thermal line widths in the kinetic energy estimations.

The root mean square (rms) velocity of the wave amplitude ($v_{\text{rms}}$) can be used to estimate the energy within a wave \citep[e.g.][]{Hollweg1981}. As such, obtaining a relation between $v_{\text{rms}}$ and the non-thermal line width ($\sigma_{\text{nt}}$) is useful for achieving a more accurate estimate of the total wave energy. Such a relation may be given by $\sigma_{\text{nt}} = \alpha v_{\text{rms}}$; however, there is some discrepancy between the value of $\alpha$ to be used, as well as a lack of any convincing justification for this chosen value. In \citet{HasslerRottman1990, BanerjeeTeriaca1998}, and \citet{DoyleBanerjee1998}, these authors computed the \Alfvenic wave energy using $\alpha\approx1/\sqrt{2}$, where the $1/\sqrt{2}$ accounts for the polarisation and direction of propagation of the wave relative to the LOS. This is the most commonly used value of $\alpha$ in estimates of the energy within an \Alfvenic wave \citep[e.g.][]{O'SheaBanerjee2005,BanerjeePerezSuarez2009,HahnLandi2012}. However, \citet{ChaeSchuhle1998} and \citet{TuMarsch1998} both suggested that $v_{\text{rms}} = \sigma_{\text{nt}}$ ($\alpha = 1$).

In order to investigate the relationship between the wave amplitude and non-thermal line width in more detail, \citet{PantVanDoorsselaere2020} (PVD2020) considered a selection of velocity drivers in a simple mathematical model. They found that for a mono-periodic linearly polarised velocity driver oscillating along the LOS, $\sigma_{\text{nt}}/v_{\text{rms}} \approx \sqrt{2}$. On the other hand, when the oscillations act in different directions (akin to the superposition of spectra of all oscillating structures along the LOS in the optically thin corona), the ratio $\sigma_{\text{nt}}/v_{\text{rms}}$ is approximately one. This value was also found when the authors used a multi-frequency driver or circularly polarised transverse oscillations. The authors confirmed their findings using forward modelling on numerical MHD simulations of transverse MHD waves in a gravitationally stratified plasma. They conclude that depending on the scenario, $\sigma_{\text{nt}}/v_{\text{rms}}>\sqrt{2}$ or $\sigma_{\text{nt}}/v_{\text{rms}}>1$; however, the ratio is never equal to $1/\sqrt{2}$ as was used in previous studies. In other words, the root mean squared wave amplitudes are never bigger than the non-thermal line widths and previous studies may have overestimated the wave energy.

In this study, we expand on the work of PVD2020 by examining the behaviour of the wave amplitudes and non-thermal line widths using a variety of more complex numerical models. Firstly, we investigate \Alfven waves in a uniform plasma and explore the effects of wave interference. Then, we consider observational signatures of transverse MHD waves propagating through a complex magnetic field. Finally, we investigate the relationship between non-thermal line widths and velocity amplitudes in simulations of heating in a coronal arcade. In Sect. \ref{sect_numerical_model}, we give an overview of these three numerical models. Then, in Sect. \ref{ntlw_vrms_section}, we explain the calculation of $v_{\text{rms}}$ and $\sigma_{\text{nt}}$ as well as analyse the results of the ratio $\sigma_{\text{nt}}/v_{\text{rms}}$ in all three models. Finally, our findings are discussed and summarised in Sect. \ref{sec_Discussion}.

\section{Numerical models} \label{sect_numerical_model}

We begin by providing a brief description of the three numerical models which we analyse in this article. All three models use the Lagrangian-remap code, Lare3D \citep{ArberLongbottom2001}, which solves the fully 3D non-ideal MHD equations in normalised form, given by

\begin{eqnarray}
    \;  \frac{D\rho}{Dt} &=& -\rho \boldnabla\cdot\vec{v},   \\
    \;  \rho\frac{D\vec{v}}{Dt} &=& \vec{j}\times\vec{B} - \boldnabla p + F_{\text{visc}},  \label{eqn_motion}\\
    \;  \rho\frac{D\epsilon}{Dt} &=& \eta j^2 - p\left(\boldnabla\cdot\vec{v}\right) + Q_{\text{visc}},  \label{energy_eqn} \\
    \; \frac{D\vec{B}}{Dt} &=& \left(\vec{B}\cdot\boldnabla\right)\vec{v} - \left(\boldnabla\cdot\vec{v}\right)\vec{B} - \boldnabla\times \left(\eta\boldnabla\times\vec{B}\right)
\end{eqnarray}

\noindent  where all variables have their usual meanings. The non-ideal terms, resistivity ($\eta$) and viscosity ($\nu$), dissipate energy from the magnetic and velocity fields, respectively. The viscosity term results in a force $F_{\text{visc}}$ in the equation of motion (\ref{eqn_motion}) and a heating term $Q_{\text{visc}}$ in the energy equation (\ref{energy_eqn}). It is the sum of the background viscosity and two small shock viscosity terms. These shock viscosities, which are present in all of the numerical models, are designed to prevent shocks and ensure numerical stability. With the exception of the shock viscosities, non-ideal terms are only included within one of the three numerical models (see Sect. \ref{arcade_model}). The effects of thermal conduction, optically thin radiation, and gravity are neglected in our simulations.

\subsection{\Alfven wave model}

\begin{figure}[t!]
\centering
\vspace{0cm}
\begin{subfigure}{0.22\textwidth}
  \centering
  \hspace{0cm}
  \makebox[0pt]{\includegraphics[width=1.\textwidth]{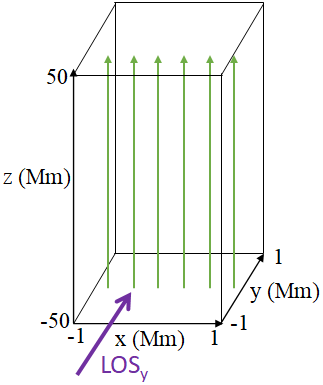}}
  \caption{}
  \label{mag_field_alfven}
\end{subfigure}%
\begin{subfigure}{0.28\textwidth}
\hspace{0cm}
  \centering
  \hspace{0cm}
\makebox[0pt]{\includegraphics[width=1.\textwidth]{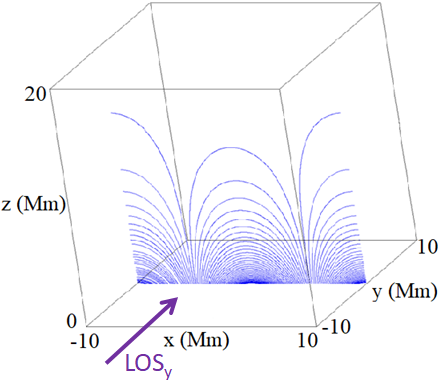}}
  \caption{}
  \label{mag_field_arcade}
\end{subfigure}
\begin{subfigure}{0.5\textwidth}
\hspace{0cm}
  \centering
  \hspace{0cm}
\makebox[0pt]{\includegraphics[width=1.\textwidth]{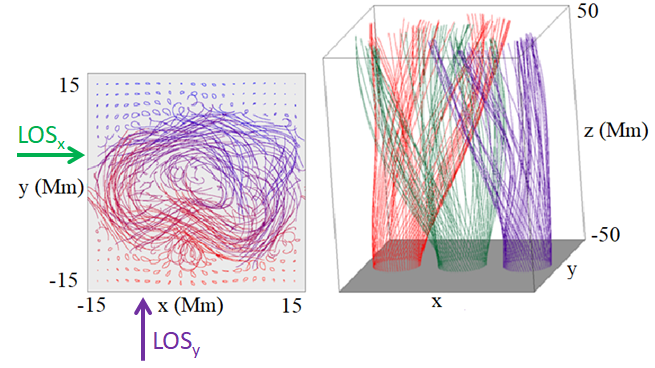}}
  \caption{}
  \label{mag_field_complex}
\end{subfigure}
\caption{Illustrations of the initial magnetic field lines in (a) the \Alfven wave model, (b) the arcade model, and (c) the complex magnetic field model. The left panel of (c) shows the projection of the field lines onto the $xy$ plane. Panels (b) and (c) were modified from \citet{HowsonDeMoortelFyfe2020}
and \citet{HowsonDeMoortel2020}, respectively. The LOS angles are denoted by $\text{LOS}_{x}$ (green) and $\text{LOS}_{y}$ (purple) when aligned with the $x$ and $y$ axes, respectively.}
\label{mag_fields}
\end{figure}

The first and simplest of our three numerical simulations is the \Alfven wave model. The setup consists of a homogeneous plasma, with a density and temperature of $1.67\times 10^{-12}$ kg $\text{m}^{-3}$ and 1.2 MK, respectively, and a uniform magnetic field (20 G) aligned with the vertical $z$ axis (see Fig. \ref{mag_field_alfven}).

\Alfven waves are driven into the system using the following condition on the bottom $z$ boundary,

\begin{equation}
    v_{y}\left(t\right) = v_{0}\sin\left(\omega t\right),
    \label{vy_driver}
\end{equation}

\noindent where the angular frequency $\omega\approx 0.42 \text{ s}^{-1}$ which results in a period of approximately 15 s. Three wave amplitudes ($v_{0}$) are considered : 12 km $\text{s}^{-1}$ (low), 24 km $\text{s}^{-1}$ (medium), and 48 km $\text{s}^{-1}$ (high). A fourth configuration is also investigated where the amplitude of the wave is 24 km $\text{s}^{-1}$, but the driver is made up of two components as follows,

\begin{equation}
 \; v_{x}\left(t\right) = v_{y}\left(t\right) = \frac{v_{0}}{\sqrt{2}} \sin\left(\omega t\right).
 \label{v_mix_driver}
\end{equation}

\noindent The LOS that we consider in the \Alfven wave model is parallel to the $y$ axis. The first wave driver (Eq. \ref{vy_driver}) acts along the LOS and the second wave driver (Eq. \ref{v_mix_driver}) oscillates at an angle of \ang{45} to the LOS. The simulations that use Eq. \ref{vy_driver} for their velocity driver are denoted by $v_{y:\chi}$ where $\chi\in\{\text{L, M, H}\}$, for low, medium, and high wave amplitudes, respectively. Finally, the fourth simulation, which uses the same amplitude as $v_{y:\text{M}}$, shall be denoted by $v_{\text{mix}}$.

\begin{figure}[t!]
\centering
\includegraphics[width=0.5\textwidth]{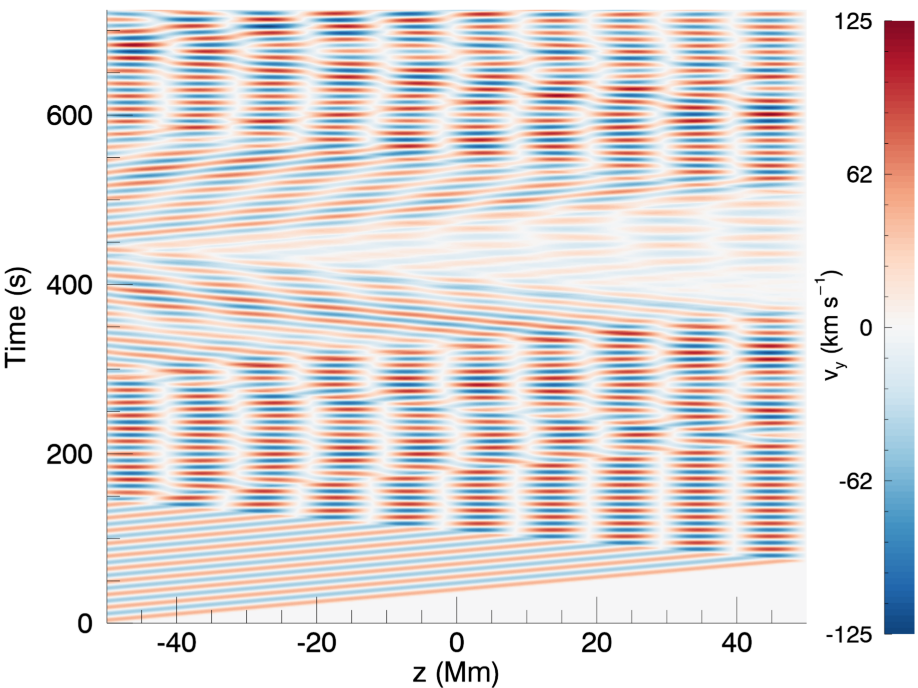}
\caption{Time-distance plot of $v_{y:\text{H}}$ (along the $z$ axis) in the \Alfven wave model. This result is independent of the $x$ and $y$ position as the model is invariant along those axes.}
\label{vy_contour}
\end{figure}

The $x$ and $y$ boundaries are periodic and the $z$ boundaries were set to have a zero gradient for all variables, with the exception of the velocity field. All components of the velocity on the $z$ boundaries are zero apart from the velocity driver on the bottom boundary, as described above. The velocity was set to zero on the top $z$ boundary to ensure that the waves are reflected here. This subsequently results in wave interference between upward and downward propagating waves. Figure \ref{vy_contour} shows a time-distance plot (along the $z$ axis) of $v_{y:\text{H}}$ (similar for $v_{y:\text{L}}$ and $v_{y:\text{M}}$). One feature which is important to the subsequent analysis of this model (see Sect. \ref{alfven_wave_analysis}) is the prevalence of nodes (e.g. at $z \approx 40$ Mm).

The computational domain has dimensions of 2 Mm $\times$ 2 Mm $\times$ 100 Mm and uses a numerical grid of 8 $\times$ 8 $\times$ 1024 cells. As this simulation is invariant in the $x$ and $y$ directions, we used a coarser grid resolution for these axes than for the $z$ axis. 

\subsection{Complex magnetic field model}

For our second model, we consider a simulation which also uses a sinusoidal boundary driver. However, in this case, the magnetic field structure is a lot more complex (complex magnetic field model). The simulation used here was previously discussed and investigated by \citet{HowsonDeMoortel2020} and subsequently forward modelled by \citet{FyfeHowsonDeMoortel2020}.

The initial magnetic field configuration in  \citet{HowsonDeMoortel2020} was derived from a simulation investigated by \citet{ReidHood2018}. In the latter article, three magnetic threads were twisted at their footpoints by rotational velocity drivers. The kink instability was triggered in the central thread which ultimately destabilised the remaining threads. The end result was a very complex magnetic field configuration which \citet{HowsonDeMoortel2020} used as their initial condition. Of the two field profiles considered in \citet{HowsonDeMoortel2020}, we only analyse the more complex state (see Fig. \ref{mag_field_complex}). The initial temperatures and densities observed within this model are approximately 1.7 MK - 4.7 MK and $1.12\times 10^{-12}\text{ kg}\text{ m}^{-3} - $$2.15\times 10^{-12}\text{ kg}\text{ m}^{-3}$, respectively.
 
Using this initial condition, the authors excited transverse waves into the numerical domain. To do this, a wave driver is imposed on the bottom $z$ boundary given by $\vec{v}\left(t\right) = \left(0,v_{y},0\right)$, where $v_y$ is defined as

 
 \begin{equation} \label{eq:complex_driver}
     v_{y}\left(t\right) = v_{0}\sin\left(\omega t\right),
 \end{equation}
 
\noindent with an amplitude and angular frequency of approximately $20\text{ km }\text{ s}^{-1}$ and $0.21\text{ s}^{-1}$, respectively. This corresponds to a period of $\tau\approx 28\text{ s}$.

As with the \Alfven wave model, the \(x\) and \(y\) boundaries are periodic while the $z$ boundaries have gradients set to zero for all variables expect for the velocity field. On the bottom $z$ boundary, the velocity driver (Eg. \ref{eq:complex_driver}) is imposed and the velocity is set to zero on the top $z$ boundary. This causes waves to reflect at the top boundary and subsequently results in wave interference from upward and downward propagating waves.

For this model, the numerical domain consists of a \({256 \times 256 \times 1024}\) grid, which covers physical dimensions of \(\text{30 Mm}\times \text{30 Mm} \times \text{100 Mm}\). However, within the forward modelling analysis in \citet{FyfeHowsonDeMoortel2020}, which we subsequently used to obtain the non-thermal line widths (see Sect. \ref{ntlw_vrms_section}), the grid used in \citet{HowsonDeMoortel2020} was spatially resampled to every fourth grid cell along $x$, $y$, and $z$. This was to reduce the computational cost and was shown to have no significant impact on the synthetic spectroscopic data. For more information on the behaviour and forward modelling of the simulation, we direct the reader to \citet{HowsonDeMoortel2020} and \citet{FyfeHowsonDeMoortel2020}, respectively.

\subsection{Arcade model} \label{arcade_model}

The last of our three numerical models considers a potential coronal arcade where a complex velocity driver is implemented. This simulation was studied by \citet{HowsonDeMoortelFyfe2020} and hence we direct the reader to this article for further information. The authors considered several numerical simulations (with different characteristic driving timescales) and they present results for ideal, resistive, and viscous regimes.

\citet{HowsonDeMoortelFyfe2020} constructed a numerical arcade within an initially homogeneous plasma with a temperature and density of approximately 1 MK and  $1.67\times 10^{-12}\text{ kg}\text{ m}^{-3}$, respectively. The arcade magnetic field has the form $\vec{B}\left(x,z\right) = \left(B_{x}, 0, B_{z}\right)$ where

\begin{eqnarray}
    \ B_x(x, z) &=& B_0\text{ cos}\left(\frac{\pi x}{L}\right)\exp{\left(\frac{-\pi z}{L}\right)},\nonumber \; \\
    \ B_z(x, z) &=& -B_0\text{ sin}\left(\frac{\pi x}{L}\right)\exp{\left(\frac{-\pi z}{L}\right)}.   \;
\end{eqnarray}

\noindent Here, $B_{0} = 100$ G and $L = 10$ Mm. Such a magnetic field is a potential field that is also invariant along the $y$ axis (see Fig. \ref{mag_field_arcade}). The domain contains $256^{3}$ grid cells with physical dimensions of $-10 \text{ Mm}\leq x,y \leq  10 \text{ Mm}$ and $0 \text{ Mm}\leq z \leq  20 \text{ Mm}$.

As mentioned previously, resistivity and viscosity are included in separate simulations, as well as including an ideal case. The non-ideal regimes allow for the dissipation of energy through the magnetic and velocity fields, respectively. A step function is used for the resistivity where it is zero for $z< 1$ Mm and $\eta_{0}$ for $z\geq 1$ Mm, where $\eta_{0}$ corresponds to a magnetic Reynolds number of $10^{4}$. The resistivity is set to zero for $z < 1$ Mm to prevent the slippage of magnetic field lines through the velocity field (with the exception of numerical slippage). Finally, the viscous simulations implement a uniform viscosity which produces a fluid Reynolds number of $10^3$.

\citet{HowsonDeMoortelFyfe2020} implemented a boundary driver which mimics this chaotic nature of photospheric motions by varying the driver in time and space. The velocity driver on $z = 0 $ Mm was created using the summation of 2D Gaussians and takes the form $\vec{v} = \left(v_x, v_y, 0\right)$ where

\begin{eqnarray}
    \ v_x &=& \sum_{i=1}^{N} v_i\cos\left(\theta_i\right)\exp\Bigg\{\frac{-\left(r-r_i\right)^2}{l_i^2}\Bigg\}\exp\Bigg\{\frac{-\left(t-t_i\right)^2}{\tau_i^2}\Bigg\}, \nonumber \; \\
    \  v_y &=& \sum_{i=1}^{N} v_i\sin\left(\theta_i\right)\exp\Bigg\{\frac{-\left(r-r_i\right)^2}{l_i^2}\Bigg\}\exp\Bigg\{\frac{-\left(t-t_i\right)^2}{\tau_i^2}\Bigg\}.  \;
\end{eqnarray}

\noindent Here, $v_{i}$, $\theta_i$, $r_i$, and $l_i$ are the peak amplitude, direction, centre, and length scale of the Gaussian components, respectively. Finally, $t_i$ and $\tau_i$ represent the time of peak amplitude and the duration of the individual Gaussian components, respectively. These quantities arise from the following statistical distributions,

\begin{eqnarray}
\  v_i \sim \mathcal{N}\left(v_{\mu}, \frac{v_{\mu}^2}{25}\right), \quad \theta_i \sim \mathcal{U}\left(0, 2\pi\right),\quad r_i\sim\mathcal{U}\left(-L,L\right),\quad \; \nonumber \\
\ l_i \sim\mathcal{N}\left(\frac{L}{4},\frac{L^2}{400}\right), \quad t_i \sim\mathcal{U}\left(t_s, t_f\right), \quad\tau_i \sim \mathcal{N}\left(\tau_{\mu}, \frac{\tau_{\mu}^2}{16}\right),\; 
\end{eqnarray}

\noindent where $\mathcal{N}\left(\mu, \sigma^2\right)$ and $\mathcal{U}\left(u_1, u_2\right)$ are the normal and uniform distributions, respectively, with mean - $\mu$, variance - $\sigma^2$, and lower and upper bounds of $u_1$ and $u_2$, respectively. The start and end time of the simulations are denoted by $t_s$ and $t_f$, respectively.

\citet{HowsonDeMoortelFyfe2020} analyse three different driving timescales and here we consider the lower and upper values $\tau_{\mu} = 15$ s and 300 s (referred to as $T_{\text{S}}$ and $T_{\text{L}}$ simulations for the short and long timescales, respectively). To allow for a comparison between the two drivers, the spatio-temporal average of the drivers' velocity were set to 1.2 km $\text{s}^{-1}$ by choosing the appropriate value for $v_{\mu}$.  The integer $N$ was chosen to be a function of the timescale $\tau_{\mu}$ to ensure that a similar number of components in the summation were active at any given time.

The boundary conditions are periodic on the $x$ and $y$ boundaries. All variable gradients are set to zero on the $z$ boundaries apart from the velocity driver, imposed on the bottom boundary. In addition, a damping layer was implemented above $z = 18$ Mm near the top of the domain. This damping layer prevents the reflection of upward flows back into the domain.


\section{Non-thermal line widths and wave amplitudes } \label{ntlw_vrms_section}

In order to investigate the relation between the non-thermal line widths and the amplitudes of the waves observed in the three models (see Sect. \ref{sect_numerical_model}), we began by measuring the wave amplitudes using the rms velocity of the waves ($v_{\text{rms}}$). As for the numerical simulations in \citet{PantMagyar2019} and PVD2020,  $v_{\text{rms}}$ was calculated as a function of height ($z$) as follows,

\begin{equation}
    v_{\text{rms}}\left(z\right) = \Bigg \langle\sqrt{\frac{\sum_{t=0}^{T-1}|\vec{v}(x,y,z,t)|^2}{T}}\Bigg\rangle_{x,y},
\end{equation}

\noindent where $T$ denotes the number of simulation output times and we averaged over the $xy$ planes for all heights. 

The synthetic specific intensity used in determining the non-thermal line width was obtained using the forward modelling code FoMo \citep{VanDoorsselaereAntolin2016}. It uses the CHIANTI atomic database \citep{DereLandi1997, LandiYoung2013} to produce optically thin EUV and UV emission lines, and it allows for different LOS angles. The emission lines and LOS angles used in our three models are summarised in the final two rows of Table \ref{model_info_table}, while Fig. \ref{mag_fields} illustrates the LOS angles. Table \ref{model_info_table} also lists the numerical cadence, exposure time, and the driver period used in the three models.

Within this article, we consider various exposure times during our simulations (see Table \ref{model_info_table} for the exact values used in each model). The exposure times were chosen such that they are not a multiple of the model's velocity driver period, with some smaller and some greater than this period. For a given exposure time, an average of the specific intensity was taken. In each case, the observing started at the beginning of the simulation. Once the specific intensity ($I_{\lambda}$) was calculated, the total intensity ($I$), Doppler shift ($\lambda_{\text{DV}}-\lambda_0$), line width ($\sigma$), and subsequently the non-thermal line width ($\sigma_{\text{nt}}$) could be calculated. This was achieved using the moments of $I_{\lambda}$ as follows,

\begin{eqnarray}
  \; I\left(x',z,t\right)  &=& \int I_{\lambda}\left(x',z,t,\lambda\right) d\lambda, \nonumber  \\
  \; \lambda_{\text{DV}}\left(x',z,t\right) &=& \frac{1}{I}\int\lambda I_{\lambda}d\lambda, \nonumber
 \\
 \; \sigma\left(x', z, t\right) &=&\sqrt{\frac{\int\left(\lambda-\lambda_{\text{DV}}\right)^2I_{\lambda}d\lambda}{I}}, \nonumber\\
 \; \sigma_{\text{nt}}\left(z\right) &=& \Bigg\langle\sqrt{\sigma_{1/e}^2-\sigma_{\text{th}}^2}\Bigg\rangle_{x',t},\label{ntlw_eq}
\end{eqnarray}

\noindent where $\sigma_{1/e}$ is the exponential line width (i.e. $\sqrt{2}\sigma$) converted into units of velocity and $\sigma_{\text{th}}$ is the thermal velocity  (Fe \rom{9}: $15.7\text{ km }\text{s}^{-1}$, Fe \rom{12}: $21.5\text{ km }\text{s}^{-1}$, and Fe \rom{16}: $27.9\text{ km }\text{s}^{-1}$) using the peak formation temperature of the emission line. As with $v_{\text{rms}}$, $\sigma_{\text{nt}}$ is also a function of the height. It was calculated using Eq. \ref{ntlw_eq} which has been used in previous work \citep[e.g.][]{TestaDePontieu2016,PantMagyar2019,PantVanDoorsselaere2020}. To denote the axes in the plane-of-sky (POS), a dash is used (e.g. $x$' denotes the horizontal axis in the POS). Since all the LOS angles considered in this article are perpendicular to the vertical axis, $z = z'$.

\begin{table}[ht!]
\caption{List of values used in the numerical models.}
    \centering
    \makebox{\begin{tabular}{|| m{1.7cm} | m{1.7cm} | m{2.1cm} | m{1.9cm} ||}
        \hline\hline
         \makecell{\textbf{Models}} & \makecell{{\Alfven} \\{wave}} & \makecell{{Complex}  \\{magnetic field}} & \makecell{{Arcade}}  \\
        \hline
         \makecell{\textbf{Numerical}\\\textbf{Cadence (s)}}  & \makecell{1.5} & \makecell{3.6} & \makecell{7.2}\\
        \hline
         \makecell{\textbf{Exposure (s)}}  & \makecell{12, 32 \& 148} & \makecell{14, 47 \& 105} & \makecell{29, 261 \& 739} \\
         \hline
         \makecell{\textbf{Driver}\\\textbf{Period (s)}}  & \makecell{15} & \makecell{28} & \makecell{15 \& 300} \\
         \hline
          \makecell{\textbf{LOS Angles}}  & \makecell{$\text{LOS}_{y}$} & \makecell{$\text{LOS}_{x}$ \& $\text{LOS}_{y}$} & \makecell{$\text{LOS}_{y}$} \\
         \hline
          \makecell{\textbf{Emission} \\\textbf{Lines}}  & \makecell{Fe \rom{9}} & \makecell{Fe \rom{12} \& Fe \rom{16}} & \makecell{Fe \rom{9}} \\
         \hline\hline
    \end{tabular}}
    \tablefoot{Each row shows the numerical cadence, exposure time, period of the drivers, LOS angles, and emission lines used in the three numerical models.The $\text{LOS}_{x}$ and $\text{LOS}_{y}$ denote the LOS angles parallel to the $x$ and $y$ axes, respectively (see Fig. \ref{mag_fields}). The emission lines have rest wavelengths ($\lambda_0$) of Fe \rom{9}: 171.073 \AA, Fe \rom{12}: 193.509 $\text{\AA,}$ and Fe \rom{16}: 335.409 \AA. The driver's period in the {arcade model} are characteristic times.}
        \label{model_info_table}
\end{table}

Using the ratio of the non-thermal line width ($\sigma_{\text{nt}}$) and the root mean squared velocity ($v_{\text{rms}}$), we now investigate the relation between these two variables in our numerical models and compare them to the ratios found in PVD2020. Table \ref{model_VP_TVD_ratios} gives an overview of our models and the most relevant (similar) models and corresponding ratios studied in PVD2020. Where our models differ from those in PVD2020, the lowest expected ratio from PVD2020 is quoted. Finally, we note that PVD2020 use exposure times equal to a multiple of their driver's period (with the exception of their multi-frequency driver), whereas in this article, we consider both exposure times which are and which are not multiples of the driving period.

\begin{table}[ht!]
\caption{Models used within this article alongside the most similar model used in PVD2020.}
    \centering
    \makebox{\begin{tabular}{|| m{2.1cm} | m{3.3cm} | m{2.3cm}  ||}
        \hline\hline
         \makecell{\textbf{Model}} & \makecell{\textbf{Corresponding} \\ \textbf{Case in PVD2020}} & \makecell{\textbf{Expected Lower}\\ \textbf{Bound on Ratio}} \\
        \hline
         \makecell{{\Alfven wave} :\\ $v_{y:\text{L}}$, $v_{y:\text{M}}$, $v_{y:\text{H}}$}  & \makecell{Mono-periodic linearly \\polarised oscillations\\along the LOS} & \makecell{$\sqrt{2}$}\\
        \hline
         \makecell{{\Alfven wave} : \\ $v_{\text{mix}}$}  & \makecell{N/A} & \makecell{1} \\
         \hline
         \makecell{{Complex} \\ {magnetic field}}  & \makecell{N/A} & \makecell{1} \\
         \hline
          \makecell{{Arcade}}  & \makecell{Multi-frequency driver} & \makecell{1} \\
         \hline\hline
    \end{tabular}}
    \tablefoot{The expected lower bound on the ratio, $\sigma_{\text{nt}}/v_{\text{rms}}$, for each model in PVD2020 is given in the final column.}
    \label{model_VP_TVD_ratios}
\end{table}

\subsection{\Alfven wave model analysis}
\label{alfven_wave_analysis}

\begin{figure}[t!]
\centering
\vspace{0cm}
\begin{subfigure}{0.48\textwidth}
  \centering
  \hspace{0cm}
  \makebox[0pt]{\includegraphics[width=1.\textwidth]{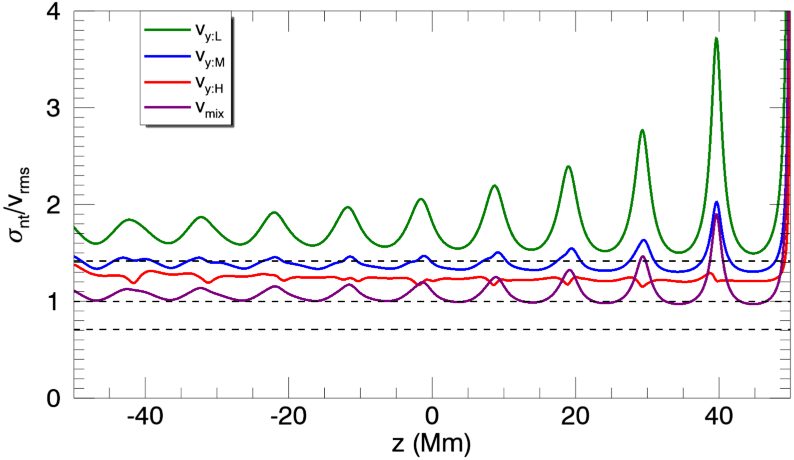}}
  \caption{12 s exposure.}
  \label{alfven_10s}
\end{subfigure}
\begin{subfigure}{0.48\textwidth}
\hspace{0cm}
  \centering
  \hspace{0cm}
\makebox[0pt]{\includegraphics[width=1.\textwidth]{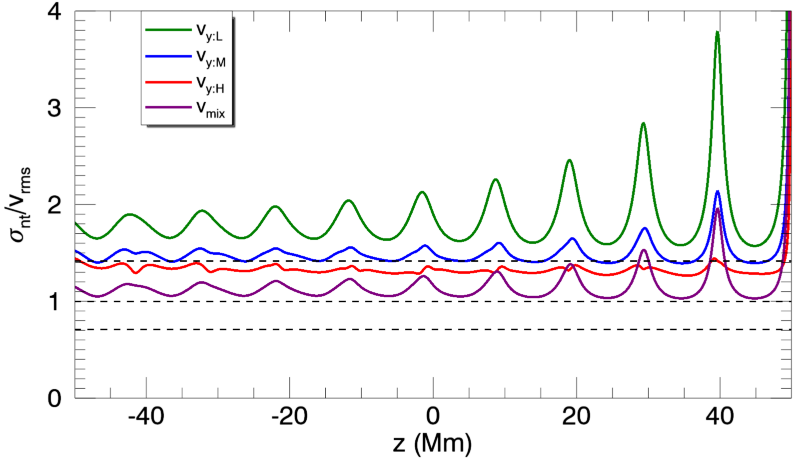}}
  \caption{32 s exposure}
  \label{alfven_30s}
\end{subfigure}
\begin{subfigure}{0.48\textwidth}
\hspace{0cm}
  \centering
  \hspace{0cm}
\makebox[0pt]{\includegraphics[width=1.\textwidth]{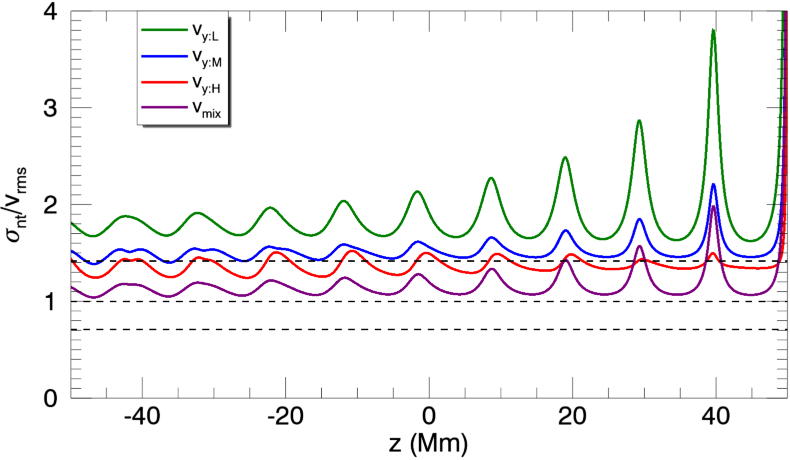}}
  \caption{148 s exposure.}
  \label{alfven_146s}
\end{subfigure}
\caption{$\sigma_{\text{nt}}/v_{\text{rms}}$ as a function of height ($z$) for the \Alfven wave model with exposure times of (a) 12 s, (b) 32 s, and (c) 148 s. The different simulations (solid lines) are $v_{y:\text{L}}$ (green), $v_{y:\text{M}}$ (blue), $v_{y:\text{H}}$ (red), and $v_{\text{mix}}$ (purple). The dashed horizontal lines, from top to bottom, are $\sqrt{2}$, 1 and $1/\sqrt{2}$.}
\label{alfven_ntlw_vrms}
\end{figure}

\begin{figure}[t!]
\centering
\includegraphics[width=0.48\textwidth]{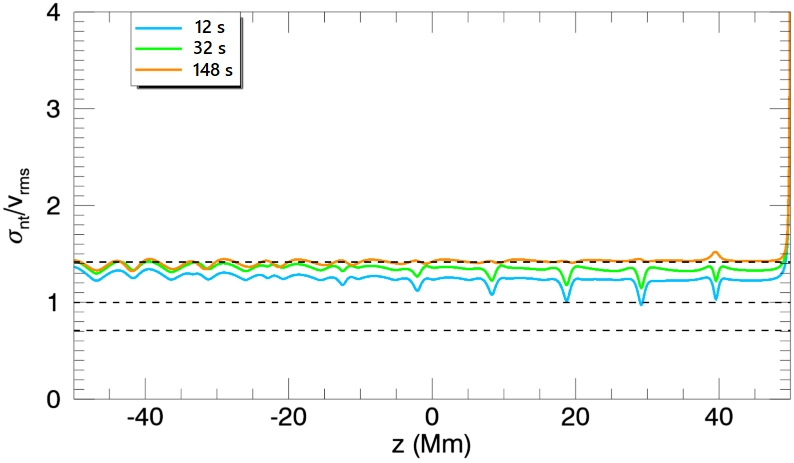}
\caption{$\sigma_{\text{nt}}/v_{\text{rms}}$ as a function of the height ($z$) for the $v_{y:\text{L}}$ simulation in the \Alfven wave model but with a plasma temperature closer to the Fe \rom{9} formation temperature. Exposure times of 12 s (light blue), 32 s (light green), and 148 s (orange) were used. The dashed horizontal lines, from top to bottom, are $\sqrt{2}$, 1 and $1/\sqrt{2}$.}
\label{alfven_ntlw_vrms_ratio_vy_low_plasma_at_fe9_temp}
\end{figure}

The ratio $\sigma_{\text{nt}}/v_{\text{rms}}$ as a function of the height for all the \Alfven model simulations and exposure times is shown in Fig. \ref{alfven_ntlw_vrms}. The first feature which clearly stands out is the presence of peaks, which correspond to nodes (see Fig. \ref{vy_contour},) and hence $v_{\text{rms}}$ is smaller on average at those heights. However, given that the velocity is on average smaller, we would also expect $\sigma_{\text{nt}}$ to decrease at these locations, leaving the ratio somewhat unaffected. This is clearly not the case and is due to our choice for the thermal line width. In real observations, the temperature in the region of interest is unknown which is why we simply selected the peak formation temperature of the emission line to represent the thermal line width. Within this current simulation, the temperature of the plasma is actually $\sim$400,000 K hotter than the Fe \rom{9} peak formation temperature. Therefore, there is an additional component within the non-thermal line width (Eq. \ref{ntlw_eq}) as the thermal line width is underestimated. The non-thermal line width is now larger than it should actually be and can be denoted by $\sigma_{\text{nt}} = \sigma_{\text{real}} + \delta$, where $\sigma_{\text{real}}$ is the true non-thermal line width and $\delta$ is the additional component due to our choice of thermal line width. As $v_{\text{rms}}$ is smaller at the altitudes which correspond to the peaks in the ratio, the ratio becomes artificially large due to the $\delta/v_{\text{rms}}$ term. To illustrate that this is indeed the case, another $v_{y:\text{L}}$ simulation was performed with a plasma temperature which is only 20,000 K above the Fe \rom{9} peak formation temperature (see Fig. \ref{alfven_ntlw_vrms_ratio_vy_low_plasma_at_fe9_temp} for the plot of its ratio versus height). As is seen in Fig. \ref{alfven_ntlw_vrms_ratio_vy_low_plasma_at_fe9_temp}, the peaks become less extreme when the plasma temperature is closer to our chosen thermal line width (the peak formation temperature of the emission line). As these peaks form for $v \approx 0$, it is unlikely that they will be seen in real observations, as some flows will always be present. However, there will be a significant additional component in the non-thermal line width in this calculation whenever $v \lesssim \delta$, which may occur frequently. This highlights the importance of selecting an appropriate thermal line width (e.g. through DEM analysis).

\begin{figure}[t!]
\centering
\includegraphics[width=0.48\textwidth]{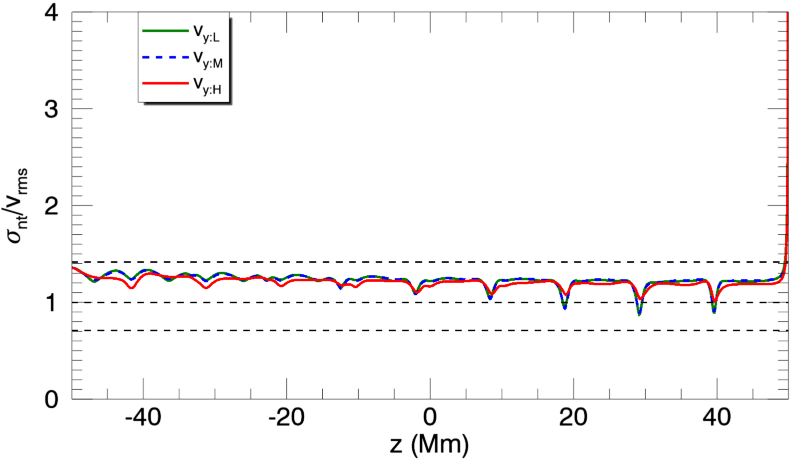}
\caption{$\sigma_{\text{nt}}/v_{\text{rms}}$ as a function of height ($z$) for the  $v_{y:\text{L}}$ (green),  $v_{y:\text{M}}$ (blue), and  $v_{y:\text{H}}$ (red) drivers in the \Alfven wave model, using a thermal line width equal to the {minimum} thermal line width in each simulation. The dashed horizontal lines, from top to bottom, are $\sqrt{2}$, 1 and $1/\sqrt{2}$.}
\label{alfven_ntlw_vrms_10s_accurate_vth}
\end{figure}

Simulations $v_{y:\text{L}}$, $v_{y:\text{M}}$, and $v_{y:\text{H}}$ show a decrease in the ratio with an increase in wave amplitude. This is the result of the additional non-thermal line width component ($\delta$) due to our estimate of the thermal line width. Similar to the behaviour of the peaks caused by the prevalence of nodes, the ratio in the $v_{y:\text{L}}$ simulation is most significantly impacted due to the smaller velocity perturbations leading to artificially larger ratios due to the $\delta/v_{\text{rms}}$ term. This term decreases with an increasing wave amplitude and we indeed see that the ratios for $v_{y:\text{L}}$, $v_{y:\text{M}}$, and $v_{y:\text{H}}$ decrease. To confirm that this behaviour is indeed caused by the additional component, $\delta$, the ratios for the  $v_{y:\text{L}}$, $v_{y:\text{M}}$, and $v_{y:\text{H}}$ simulations were calculated using the minimum thermal line width present in each simulation ($18.4 \text{ km}\text{ s}^{-1}$, $18.3 \text{ km}\text{ s}^{-1}$, and $17.6 \text{ km}\text{ s}^{-1}$, respectively). The newly calculated ratios are illustrated in Fig. \ref{alfven_ntlw_vrms_10s_accurate_vth} which uses an exposure time of 12 s and is hence comparable with Fig. \ref{alfven_10s} which uses the peak formation temperature as the thermal line width ($15.7\text{ km}\text{ s}^{-1}$: less than the thermal line widths in Fig. \ref{alfven_ntlw_vrms_10s_accurate_vth}). When the additional component of the thermal line width is reduced, by changing the thermal line width from the peak formation temperature to the minimum thermal line width, the three simulations produce similar ratios, all of which are below the ratio of $\sqrt{2}$ given in PVD2020 (see Table \ref{model_VP_TVD_ratios}). This clearly illustrates the importance of an appropriate thermal line width. In real observations, it is not always possible to determine the exact temperature of the plasma. The difference in the thermal line widths (i.e. the minimum thermal line width present and the thermal line width at the peak formation temperature) is approximately $2-3 \text{ km}\text{ s}^{-1}$ and only has a noticeable effect on the $v_{y:\text{L}}$ and $v_{y:\text{M}}$ simulations; hence the approximate additional component of the thermal line width is 10\%-23\%  of the velocity driver's amplitude for these two simulations. This suggests that any additional component in the thermal line width greater than 10\% of the velocity driver's amplitude will result in artificially large ratios.

\begin{figure}[t!]
\centering
\vspace{0cm}
\begin{subfigure}{0.48\textwidth}
  \centering
  \hspace{0cm}
  \makebox[0pt]{\includegraphics[width=1.\textwidth]{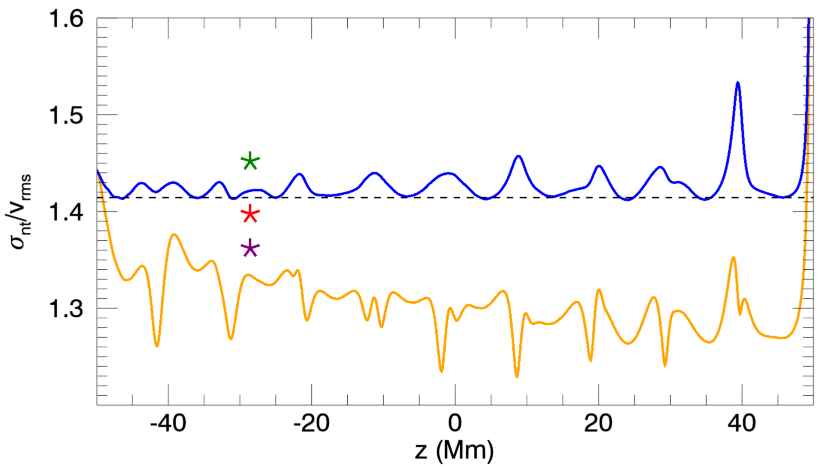}}
  \caption{$\sigma_{\text{nt}}/v_{\text{rms}}$ as a function of height ($z$) for the $v_{y:\text{H}}$ simulation in the {\Alfven wave model}. The orange line has an exposure time equal to the period of the driver, and the blue line has an `infinite' exposure time (full length of the simulation). The dashed horizontal line is at $\sqrt{2}$. The asterisks are evaluated at $(x',z) = (0.1,-28.6)$ Mm during a time frame before (green and red) and after (purple) wave reflection off the top boundary, with exposure times equal to one period (15 s: green and purple) and one and a half periods (22.5 s: red) of the velocity driver.}
  \label{alfven_vy_high_tests}
\end{subfigure}
\begin{subfigure}{0.48\textwidth}
\hspace{0cm}
  \centering
  \hspace{0cm}
\makebox[0pt]{\includegraphics[width=1.\textwidth]{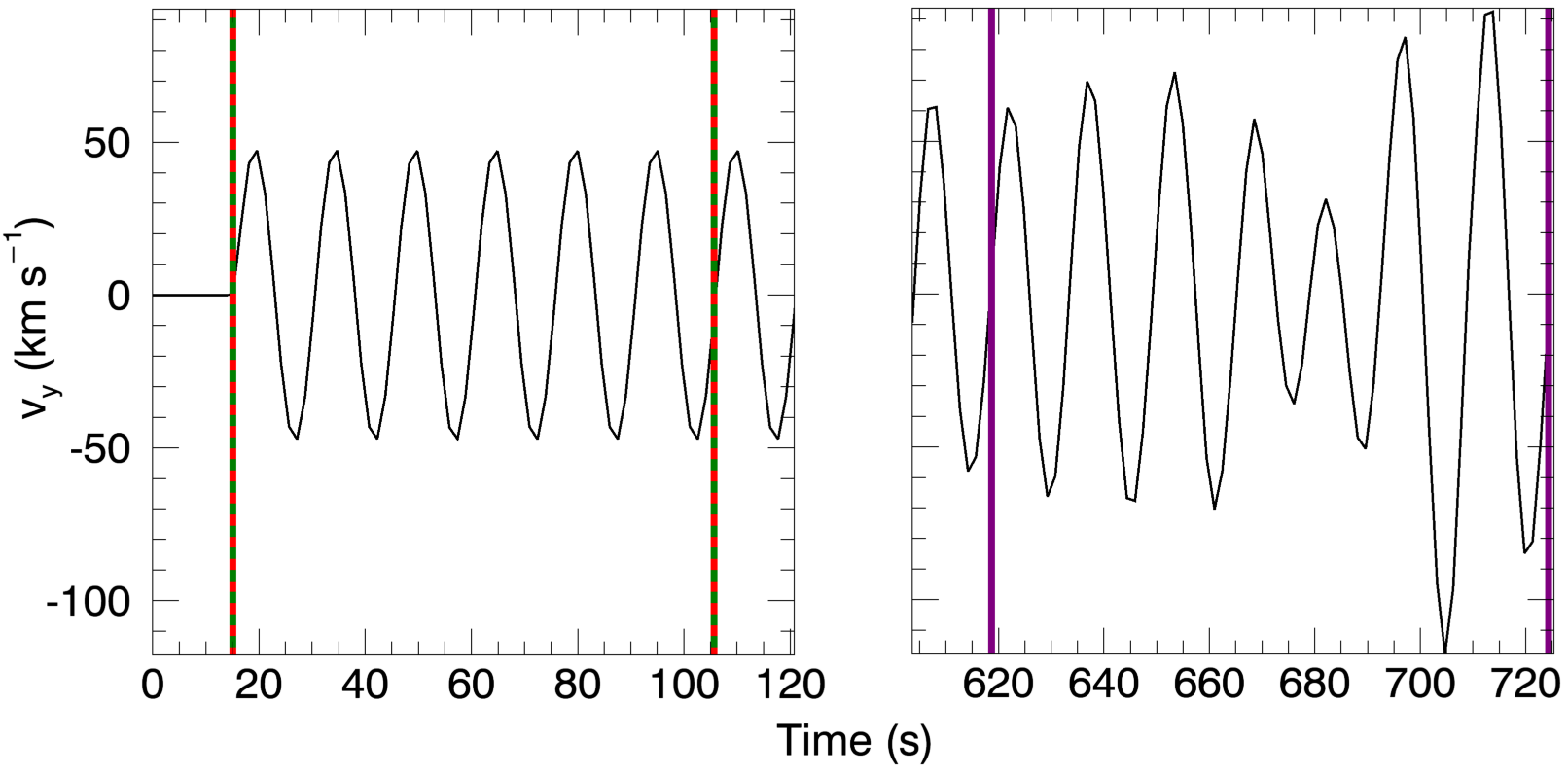}}
  \caption{Time frames for the asterisks described in (a), depicted by the vertical bars (corresponding colours) on top of $v_y$ at the point  $(x',y',z) = (0.1,0.1,-28.6)$ Mm as a function of the time (black line). Left and right-hand panels are from approximately 0-120 s and 605-725 s, respectively.}
  \label{alfven_vy_hight_test_explained}
\end{subfigure}
\caption{Analysis of $\sigma_{\text{nt}}/v_{\text{rms}}$.}
\end{figure}

\begin{figure}[ht!]
\centering
\includegraphics[width=0.48\textwidth]{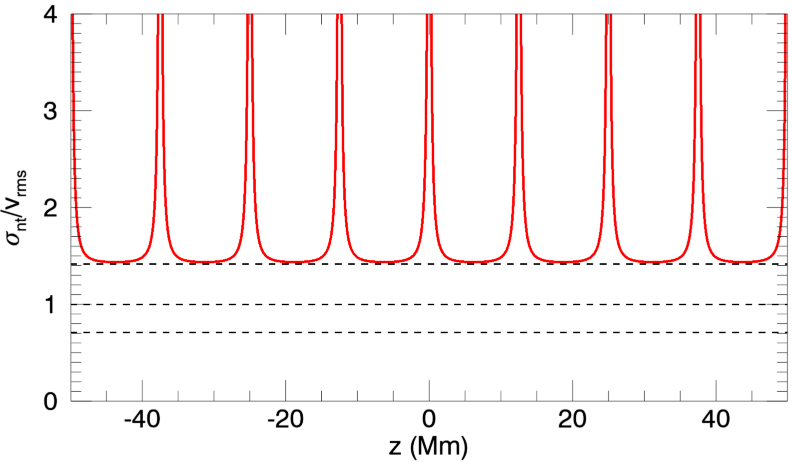}
\caption{$\sigma_{\text{nt}}/v_{\text{rms}}$ as a function of the height ($z$) for the standing wave model (same wave amplitude as $v_{y:\text{H}}$).The red line has an exposure time equal to the period of the driver. The dashed horizontal lines, from top to bottom, are $\sqrt{2}$, 1 and $1/\sqrt{2}$.}
\label{alfven_standing_wave}
\end{figure}

The closest comparison to the simulations $v_{y:\text{L}}$, $v_{y:\text{M}}$, and $v_{y:\text{H}}$ are the mono-periodic linearly polarised oscillations along the LOS in PVD2020; hence we would expect the ratio to be greater than or equal to $\sqrt{2}$ (see Table \ref{model_VP_TVD_ratios}) if the spectra are averaged over one or multiple periods of the driver. However, in our study, we have chosen exposure times that are not an exact multiple of the velocity driver's period. As seen from Fig. \ref{alfven_ntlw_vrms}, $v_{y:\text{H}}$ (red line) is the only one of the three simulations which does not satisfy $\sigma_{\text{nt}}/v_{\text{rms}} > \sqrt{2}$ (with $v_{y:\text{M}}$ below the criterion for smaller exposure times). This effect, as discussed previously, is the consequence of the additional component in the non-thermal line width ($\delta$) and all three simulations are in fact below this ratio when the `minimum' thermal line width is used (see Fig. \ref{alfven_ntlw_vrms_10s_accurate_vth}). To allow for a comparison between our simulations and the equivalent in PVD2020, we re-calculated the ratios with an exposure time equal to the driver's period, while still using the peak formation temperature as the thermal line width. When the new exposure time is applied to the $v_{y:\text{H}}$ simulation, the ratio becomes the orange line in Fig. \ref{alfven_vy_high_tests}. There is little difference in the ratio between an exposure time equal and not equal to a multiple of the driver's period (see the orange line in Fig. \ref{alfven_vy_high_tests} and the red line in the first panel of Fig. \ref{alfven_ntlw_vrms}, respectively). As a comparison, we also analyse the ratio from a simulation of a standing wave (no wave interference present) which has the same amplitude as $v_{y:\text{H}}$. When an exposure time equal to the period of the driver is used, we satisfy the $\sqrt{2}$ criterion (see Fig. \ref{alfven_standing_wave}), unlike the ratio in the $v_{y:\text{H}}$ simulation.

\begin{figure}[t!]
\centering
\includegraphics[width=0.45\textwidth]{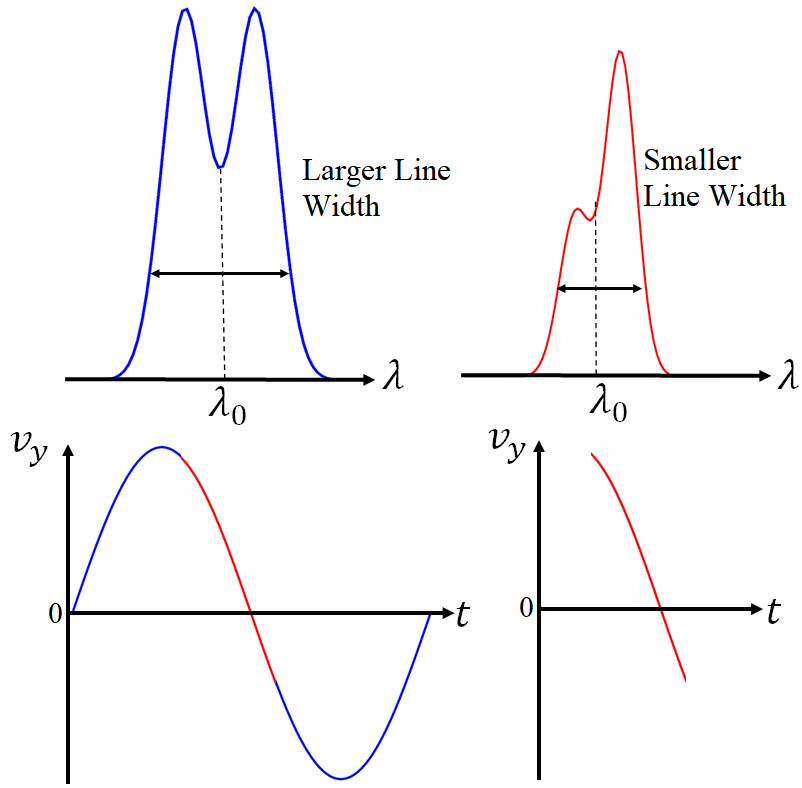}
\caption{Example of under-sampling a wave period. The bottom row is the wave present over two different exposure times. The red wave (right) is a subset of the blue wave (left) where the blue wave has an exposure time equal to the period of the driver. The top row depicts the specific intensity over the two different exposure times.}
\label{under_sampling_illustration}
\end{figure}

To explain this result, we considered the effect due to the exposure time and the effect due to the wave interference. Firstly, we shall consider the effects of the exposure time with no wave interference. At a single point in the POS ($(x',z)=(0.1,-28.6)$ Mm), a time frame was examined during the $v_{y:\text{H}}$ simulation before the first reflected wave front reached this altitude ($z = -28.6$ Mm). The ratio calculation was evaluated for two different exposure times. One exposure time is equal to the driver period (15 s) and one is not (22.5 s). These are denoted by the green and red asterisks in Fig. \ref{alfven_vy_high_tests}, respectively. Figure \ref{alfven_vy_hight_test_explained} illustrates the wave behaviour and time frames over which the asterisks in Fig. \ref{alfven_vy_high_tests} were calculated. The left-hand panel shows a time frame before wave interference and the right-hand panel shows a time frame during interference. When no wave interference is present, we see a lower ratio when the exposure time is not a multiple of the driver. Since $v_{\text{rms}}$ has no influence on the difference between the ratios, as it is the same for these two cases, we focus on the non-thermal line width. When anti-parallel flows are present along the LOS  within an exposure time, the specific intensity becomes double peaked. Whether these peaks are symmetric about  $\lambda_{\text{DV}}$, is dependent on the exposure time used. If the exposure time is equal to a multiple of the driver's period, then the specific intensity is symmetric. Conversely, if the exposure time is not a multiple of the driver's period, then under-sampling a wave period causes asymmetry. As the line width is controlled by the variation in the velocity profile along the LOS, this under-sampling can result in a decrease in the total line width. For example, this happens if the extrema in the velocity profile do not occur during the exposure time. Figure \ref{under_sampling_illustration} depicts such an example by illustrating the resultant specific intensity (top row) from a wave (bottom row) equal to the period of the driver (left column) and a wave with a period less than the driver (right column). As a result of the under-sampling, the ratio $\sigma_{\text{nt}}/v_{\text{rms}}$ typically decreases when exposure times are not a multiple of the driver's period.

 However, when considering the full simulation and using an exposure time equal to the period of the driver, the ratio does not increase as described above (readers can compare the red line in Fig. \ref{alfven_ntlw_vrms} to the orange line in Fig. \ref{alfven_vy_high_tests}, i.e. the ratio without and with an exposure time equal to the driver's period, respectively). This is due to the presence of wave interference. To analyse this, two cases are considered, one, as before, which uses a time frame before any wave interference is present (green asterisk in Fig. \ref{alfven_vy_high_tests}) and another which uses a time frame during wave interference (purple asterisk in Fig. \ref{alfven_vy_high_tests}) in the $v_{y:\text{H}}$ simulation. Both cases use an exposure time equal to the period of the wave's driver. For the purple asterisk, asymmetric specific intensities are obtained, but in this case, this is due to the wave interference. As a result of the asymmetry, the ratio in Fig. \ref{alfven_vy_high_tests} also decreases. Another contributing factor is the change in $v_{\text{rms}}$ due to the unequal wave amplitudes as a result of wave interference (readers are encouraged to compare the wave profiles in the two panels of Fig. \ref{alfven_vy_high_tests}). These two factors explain why the criterion for the ratio is satisfied in the standing wave case (Fig. \ref{alfven_standing_wave}) and not in the $v_{y:\text{H}}$ simulation (orange line in Fig. \ref{alfven_vy_high_tests}) even though an exposure time equal to the period of the driver was used.

One circumstance for which the $\sqrt{2}$ criterion is attained for the high amplitude \Alfven wave simulation when the peak formation temperature is the thermal line width (see red lines in Fig. \ref{alfven_ntlw_vrms}) is when an `infinite' exposure time is considered (see blue line in Fig. \ref{alfven_vy_high_tests}), that is the exposure time is equal to the length of the simulation. In fact, when investigating the ratio for numerous exposure times, it was found that any exposure time greater than approximately 220 s sufficed. Since the frequency of the imposed driving does not match the natural, fundamental frequency (or one of the higher harmonics), once the wave reflects off the boundaries, a beating behaviour occurs. This leads to the presence of longer periodicities in the domain than the period of the driver. As the wave amplitude changes over short times, in order to obtain a representative view of the wave behaviour, we need the exposure time to be greater than the beating period.  In other words, in order to obtain the ratio found in PVD2020, we need larger exposure times (or more periods). Within observations it is unlikely that the footpoint motions are monoperiodic, as is the case within this model, and hence this result may occur to a lesser extent within the corona.



We now consider the ratio generated from the $v_{\text{mix}}$ simulation (see purple curves in Fig. \ref{alfven_ntlw_vrms}). Firstly, the ratio is approximately a factor of $\sqrt{2}$ less than the ratio achieved in the $v_{y:\text{M}}$ simulation regardless of the thermal line width used (i.e. either the peak formation temperature or the more accurate thermal line width). This is due to the difference in the alignment of both the drivers in comparison to the $y$ axis: The component of the velocity along the LOS is a factor of $\sqrt{2}$ smaller in the $v_{\text{mix}}$ simulation, but $v_{\text{rms}}$ is the same in both cases. From the scenarios considered in PVD2020, $\sigma_{\text{nt}}>v_{\text{rms}}$ is a lower bound on the ratio (see Table \ref{model_VP_TVD_ratios}). This is indeed satisfied within the $v_{\text{mix}}$ simulation; however, when a more accurate thermal line width was used, it was determined that the ratio did not satisfy $\sigma_{\text{nt}}>v_{\text{rms}}$. We did not apply the condition $\sigma_{\text{nt}}>\sqrt{2}v_{\text{rms}}$ as the LOS is no longer aligned with the direction of oscillation. Irrespective of whether the simulations meet the conditions presented in PVD2020, none of the ratios -- including $v_{y:\text{L}}$, $v_{y:\text{M}}$, and $v_{y:\text{H}}$ -- reach $1/\sqrt{2}$, that is~the ratio which has been used over the past decade in several studies. This is also the case when a more accurate thermal line width is implemented.

\subsection{Complex magnetic field model analysis}

\begin{figure}[t!]
\centering
\vspace{0cm}
\begin{subfigure}{0.4495\textwidth}
\hspace{0cm}
  \centering
  \hspace{0cm}
\makebox[0pt]{\includegraphics[width=1.\textwidth]{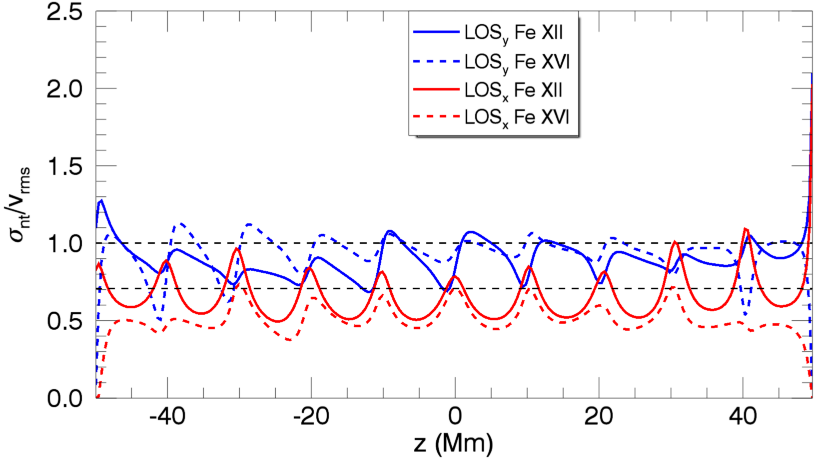}}
  \caption{14 s exposure}
  \label{complex_field_30s2}
\end{subfigure}
\begin{subfigure}{0.4495\textwidth}
\hspace{0cm}
  \centering
  \hspace{0cm}
\makebox[0pt]{\includegraphics[width=1.\textwidth]{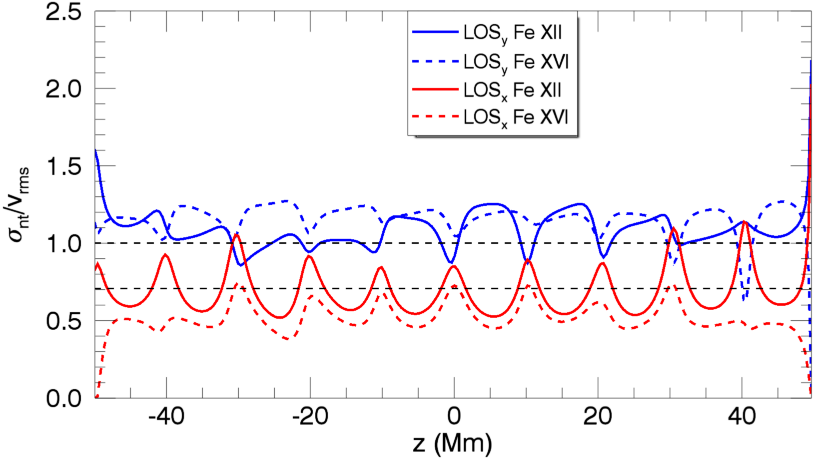}}
  \caption{47 s exposure.}
  \label{complex_field_43s2}
\end{subfigure}
\begin{subfigure}{0.4495\textwidth}
\hspace{0cm}
  \centering
  \hspace{0cm}
\makebox[0pt]{\includegraphics[width=1.\textwidth]{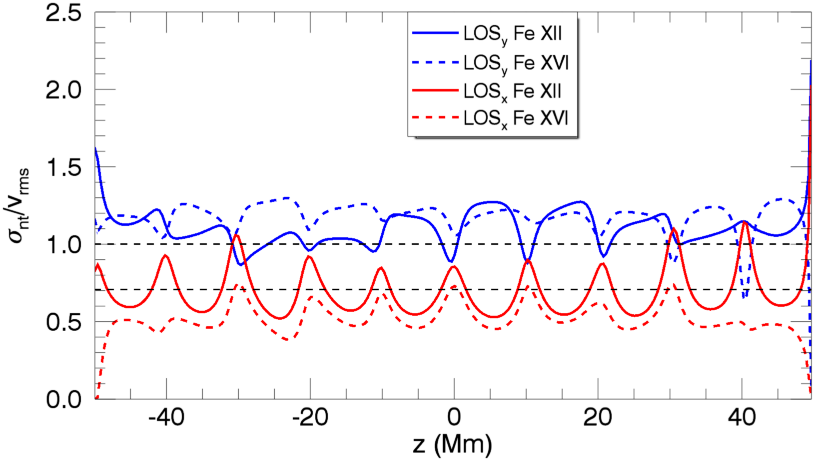}}
  \caption{105 s exposure.}
  \label{complex_field_101s2}
\end{subfigure}
\caption{$\sigma_{\text{nt}}/v_{\text{rms}}$ as a function of the height ($z$) for the {complex magnetic field model} with exposure times of (a) 14 s, (b) 47 s, and (c) 105 s. The different LOS angles are $\text{LOS}_y$ (blue) and  $\text{LOS}_x$ (red) and the emission lines are Fe \rom{12} (solid lines) and Fe \rom{16} (dashed lines). The dashed horizontal lines, from top to bottom, are 1 and $1/\sqrt{2}$.}
\label{complex_ntlw_vrms2}
\end{figure}

Within the complex magnetic field model analysis, some locations in the domain contained plasma at temperatures lower\ the peak formation temperature of Fe \rom{16} and hence Eq. \ref{ntlw_eq} had no real solutions. Two approaches were taken. The first approach was to neglect these locations in the averaging. The second approach was to set the non-thermal line widths at those problematic points to zero. These two approaches did not differ significantly; hence we have only included the latter in Fig. \ref{complex_ntlw_vrms2}, which illustrates $\sigma_{\text{nt}}/v_{\text{rms}}$ as a function of the height during the {complex magnetic field} model simulations.

Even though the driver is mono-periodic and linearly polarised along $\text{LOS}_{y}$, comparing it to the simulation with the equivalent driver in PVD2020 and, hence, using a ratio threshold of $\sqrt{2}$ is not an appropriate comparison. This is firstly because we considered $\text{LOS}_x$, which is not aligned along the direction of the driver, and secondly \citet{HowsonDeMoortel2020} and \citet{FyfeHowsonDeMoortel2020} have shown that the polarisation of the waves changes from strictly $v_y$ at the driver to also containing a $v_x$ component throughout the rest of the 3D domain. Therefore, this simulation must be compared to the lowest threshold PVD2020 present; hence examining the ratios with respect to the threshold $\sigma_{\text{nt}}/v_{\text{rms}}> 1$ is used here  (see Table \ref{model_VP_TVD_ratios}). From Fig. \ref{complex_ntlw_vrms2}, it is clear that this threshold is not always satisfied.

Firstly, observations along $\text{LOS}_y$ only achieve the threshold of one for larger exposure times, whereas $\text{LOS}_x$ fails to attain this target entirely. Increasing the exposure time does increase the non-thermal line width and hence the ratio $\sigma_{\text{nt}}/v_{\text{rms}}$  (as with the {\Alfven wave model}). However, even when an `infinite' exposure time was implemented, there was little difference between that and Fig. \ref{complex_field_101s2} which has an exposure time of 105 s. We see that observations along $\text{LOS}_x$ never attain the threshold. This is a consequence of the non-thermal line width since $v_{\text{rms}}$ is the same for both LOS angles. In Fig. 20 of \citet{FyfeHowsonDeMoortel2020}, it is shown that the mean magnitude of $v_{y}$ is greater than that of $v_{x}$. Hence, the non-thermal line width is smaller along $\text{LOS}_x$ compared to $\text{LOS}_y$. In essence, $\text{LOS}_x$ is not observing the dominant component of the velocity field ($v_{y}$) even though it is included in $v_{\text{rms}}$, and hence the ratio is less than one. The same but less extreme effect is causing the ratio to decrease for $\text{LOS}_y$. Indeed, it is below one for smaller exposure times. This effect not only explains why $\text{LOS}_x$ does not attain the threshold, but also reveals why there is a difference between the two LOS angles.

Two more factors which may influence the ratio are the complexity of the field and the presence of wave interference. In the {\Alfven wave model}, we showed that wave interference decreased the ratio as a result of the asymmetric specific intensities. In a similar way, we see asymmetric line profiles for the complex field due to wave interference and phase mixing along the LOS \citep{HowsonDeMoortel2020} and, hence, a reduction in the ratio. Here, we only consider exposure times that are not equal to a multiple of the drivers' period; however, we know from the {\Alfven wave model} that even with an exposure time equal to the period of the driver, the presence of wave interference still decreases the ratio below the anticipated threshold of one.

\begin{figure}[t!]
\centering
\includegraphics[width=0.4495\textwidth]{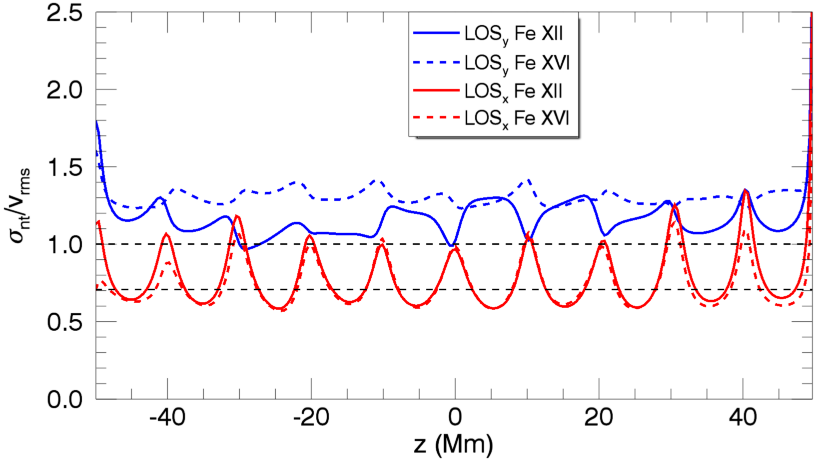}
\caption{$\sigma_{\text{nt}}/v_{\text{rms}}$ as a function of the height ($z$) for the {complex magnetic field model} with an exposure time of 105 s. The different LOS angles are $\text{LOS}_y$ (blue) and  $\text{LOS}_x$ (red) and the emission lines are Fe \rom{12} (solid lines) and Fe \rom{16} (dashed lines); however, the minimum formation temperature was used rather than the peak formation temperature. The dashed horizontal lines, from top to bottom, are 1 and $1/\sqrt{2}$.}
\label{complex_ntlw_vrms_min_formation_temp}
\end{figure}

As seen in Fig. \ref{complex_ntlw_vrms2}, not only does $\text{LOS}_{x}$ not reach the threshold of one, but it also sits on (or below, dependent on the emission line) the ratio $1/\sqrt{2}$. Previously PVD2020 found that this was not attainable in their model. We do see an increase in the ratio (see Fig. \ref{complex_ntlw_vrms_min_formation_temp}) when thermal line widths approximately equal to the minimum formation temperature of the ions are used (Fe \rom{12}: 18.4 $\text{km}\text{ s}^-1$ and Fe \rom{16}: 22.9 $\text{km}\text{ s}^-1$). However, even in this case, $\text{LOS}_{x}$ still crosses the ratio of $1/\sqrt{2}$ and hence the root mean squared wave amplitude is greater than the non-thermal line width, contrary to  PVD2020. The discrepancy between the {complex magnetic field model} and the findings of PVD2020 lies in the complexity of the models and the LOS angles. Indeed, when the LOS is parallel to the velocity driver, the two factors which produce a decrease in the ratio are the presence of wave interference and the changing polarisation of the wave. These two factors, combined with a LOS perpendicular to the velocity driver, generated ratios even less than those aligned with the driver. All of these are factors which are not present in PVD2020. And finally, as in the {\Alfven wave model}, there is an additional component in the non-thermal line width ($\delta$) due to underestimating the true thermal line width by using the peak formation temperature. This means that if a more accurate thermal line width is used, these ratios will be even smaller and a larger discrepancy will be present between this model and PVD2020.

\subsection{Arcade model analysis}

\begin{figure}[t!]
\centering
\includegraphics[width=0.4495\textwidth]{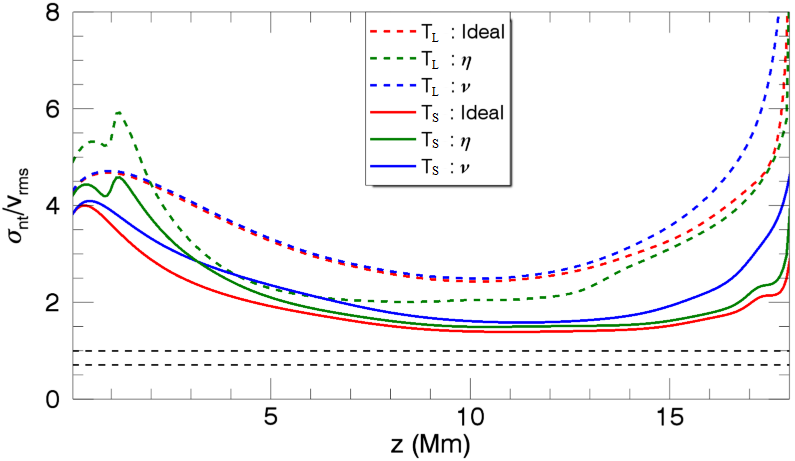}
\caption{$\sigma_{\text{nt}}/v_{\text{rms}}$ as a function of height ($z$) for the {arcade model} with an exposure times of 29 s (similar for 261 s and 739 s). The different driving timescales are $T_{\text{L}}$ (coloured dashed lines) and $T_{\text{S}}$ (solid lines) with the ideal (red), resistive (green), and viscous (blue) regimes. The dashed horizontal lines, from top to bottom, are 1 and $1/\sqrt{2}$.}
\label{arcade_ntlw_vrms}
\end{figure}

The final model examined within this article is the {arcade model}. Fig. \ref{arcade_ntlw_vrms} shows the ratio $\sigma_{\text{nt}}/v_{\text{rms}}$ as a function of the height ($z$) for various simulations using the 29 s exposure time. The 261 s and 739 s exposure times generated very similar ratios and hence have been neglected in the figure. Due to the damping layer close to the top $z$ boundary and the effect it has on $v_{\text{rms}}$, we neglected $z > 18$ Mm. The threshold of one, from the multi-frequency driver simulation in PVD2020, is used for comparison with this current model (see Table \ref{model_VP_TVD_ratios}). From Fig. \ref{arcade_ntlw_vrms}, it is clear that all regimes (ideal, resistive, and viscous) and driving timescales (short and long) are above the threshold. However, we need to err on the side of caution with this analysis. The thermal line width is underestimated by just under $2 \text{ km}\text{ s}^{-1}$, as we used the peak formation temperature of the ion rather than the temperature of the simulation. This would not be an issue if the velocity perturbations present were significantly larger; however, the average of $v_{\text{rms}}$ is approximately $1.5 \text{ km}\text{ s}^{-1}$. As with the peaks and ratios seen in the {\Alfven wave model}, the additional component in the non-thermal line width ($\delta$), alongside the low velocity perturbations, causes the ratio to be larger than it should actually be. To confirm this, one of the simulations was analysed again with the velocity field artificially increased to be twenty times greater than the original simulation. In this case, the ratio decreased to a value of about $1/\sqrt{2}$. Therefore, further analysis of this model is somewhat unreliable as the term $\delta/v_{\text{rms}}$ is dominating the behaviour of the ratio.

In \citet{HowsonDeMoortelFyfe2020}, the authors demonstrate that more heat is generated in the $T_{\text{L}}$ simulation than in the $T_{\text{S}}$ simulation. Bearing this in mind, the increase in the ratio from the $T_{\text{S}}$ simulation to the $T_{\text{L}}$ simulation, seen in Fig. \ref{arcade_ntlw_vrms}, may be due to the increased heating in the $T_{\text{L}}$ simulation, alongside the constant thermal line width used for both simulations. More specifically, there is a larger additional component in the non-thermal line width ($\delta$) in the $T_{\text{L}}$ simulation than in the $T_{\text{S}}$ simulation.

Finally, when comparing the results for different exposure times, unlike the {\Alfven wave model} and the {complex magnetic field model}, there is very little difference between the ratios. This, however, is difficult to examine as it is most likely due to the overshadowing of the large additional component in the non-thermal line width.

\section{Discussion and conclusions} \label{sec_Discussion}

In this paper, we have expanded on the work of \citet{PantVanDoorsselaere2020}, where the authors examine the relation between the root mean squared wave amplitudes ($v_{\text{rms}}$) and the non-thermal line widths ($\sigma_{\text{nt}}$). The ratio $\sigma_{\text{nt}}/v_{\text{rms}}$ was frequently used to estimate observed wave energies. However, PVD2020 claim that the value of this ratio is incorrect and that previous wave energies have possibly been overestimated. In this article, we look at more complex MHD models than the ones investigated in PVD2020 in order to determine if their claim still holds. 

To be able to estimate the non-thermal line width from observed line profiles, it is necessary to first establish the thermal component of the line width. To mimic the information available in actual observations, we based the thermal line width in our study on the peak formation temperature of the emission line (unless otherwise stated for comparative purposes), rather than the actual temperatures in the 3D simulation domains. However, when the temperature in the simulation domain is larger than the peak formation temperature, this can affect the reliability of the ratio $\sigma_{\text{nt}}/v_{\text{rms}}$.  Indeed, by writing the non-thermal line width as $\sigma_{\text{nt}}=\sigma_{\text{real}} + \delta$, where $\sigma_{\text{real}}$ is the `true' non-thermal line width and $\delta$ is the additional component due to underestimating the thermal line width, it is clear that the ratio $\sigma_{\text{nt}}/v_{\text{rms}}$ becomes larger than it should actually be. When velocities in the domain are small, this additional component in the non-thermal line width can dominate the ratio. By comparing simulations, we deduced that the $\sigma_{\text{nt}}/v_{\text{rms}}$ ratio becomes unreliable in locations where the additional component in the thermal line width is greater than about 10\% of the velocities. Another scenario is when the plasma temperatures are less than our chosen thermal line width. When this is the case, the non-thermal line width (see Eq. \ref{ntlw_eq}) has no real solution and hence the analysis breaks down.

As well as the thermal line width, the choice of exposure time was also found to affect the ratio $\sigma_{\text{nt}}/v_{\text{rms}}$. Again, to reflect actual observations, we chose to make the exposure times independent of the period of the drivers in our simulations. In other words, the exposure time was not chosen to be an exact multiple of the period. It was found that when a non-integer multiple of the driver's period was used as the exposure time, the ratio would decrease in comparison to an exposure time equal to a multiple of the period of the driver. This was due to the under-sampling of wave periods when the exposure time did not equal a multiple of the driver's period, resulting in smaller non-thermal line widths (see Fig. \ref{under_sampling_illustration} for an example of under-sampling). One method, however, that was found to increase the ratio was to use larger exposure times. This increased the ratio in some of the simpler simulations (i.e. {\Alfven wave model}), such that the ratio coincided with that in PVD2020, when previously they did not with smaller non-integer multiples of the driver's period as the exposure time. Due to the multi-frequency nature of the corona, this exposure time result may not emerge in observations as our simulations contain monoperiodic drivers.


Another influential factor in the ratio $\sigma_{\text{nt}}/v_{\text{rms}}$ is the presence of wave interference. It was found to decrease the ratio when comparing a simple \Alfven wave model without and with the reflection of the wave off the top boundary (i.e. generating wave interference).

Within the {complex magnetic field model}, both the exposure time and the wave interference played important roles in reducing the ratio between the non-thermal line width and the root mean squared wave amplitudes. In addition, the LOS angle was also found to play a critical role. Two LOS angles were considered in this model, one parallel ($\text{LOS}_{y}$) and one perpendicular ($\text{LOS}_x$) to the velocity driver ($v_y$) on the bottom boundary. Throughout the simulation, the mean magnitude of $v_y$ is greater than that of $v_x$  \citep[see Fig.~20 of][]{FyfeHowsonDeMoortel2020}. Hence, $\text{LOS}_x$ is not observing the dominant component of the velocity field, but it is included in the $v_{\text{rms}}$ calculation. This resulted in not only $\text{LOS}_x$ producing ratios less than $\text{LOS}_y$, but also generating a ratio which is less than the one predicted in PVD2020 ($\sigma_{\text{nt}}/v_{\text{rms}}>1$). For $\text{LOS}_y$, the ratio only reaches one for larger exposure times and is less than one for smaller exposure times.

Our models use a static background with waves driven using mono- or multi-periodic drivers. This setup is simplistic in comparison to the corona's more dynamic behaviour, where the background is not necessarily time-independent and waves can be turbulently driven. Although we consider spatial complexity in our {complex magnetic field model,} the complexity of the field is still time-independent. If temporal variations to the background on timescales similar to the waves were also present, identifying waves might no longer be possible. For example, \citet{GoossensArregui2019} show that spatial complexity mixes the properties of the MHD waves; this also holds when temporal variations are present. However, we found that using the ratio $\sigma_{\text{nt}}/v_{\text{rms}}$ to estimate wave energies is not a robust approach and this conclusion equally holds if short-timescale temporal variations in a dynamically changing corona are present.


Our analysis has highlighted several key issues which need to be taken into account when estimating wave energy  budgets from observations. For example, it is important that an appropriate thermal line width is selected and this is not necessarily the formation temperature of the emission line under investigation. One method of obtaining a more accurate thermal line width is through DEM analysis. From the numerical models presented in this article, the average value of $\sigma_{\text{nt}}/v_{\text{rms}} = 1.7$. Although this average satisfies the findings of PVD2020 (i.e. either $\sigma_{\text{nt}}/v_{\text{rms}}>\sqrt{2}$ or $\sigma_{\text{nt}}/v_{\text{rms}}>1$ dependent on the scenario), we do find that the ratio for the different models ranges from 0.38 to 5.92 (neglecting the values caused by boundary conditions). Overall, the ratio is highly dependent on a number of factors (e.g. LOS angles, magnitude of the velocity perturbations, presence of wave interference, and the length of the exposure time) and hence it is not possible to identify a single value for the $\sigma_{\text{nt}}/v_{\text{rms}}$ ratio.

\vspace{1cm}
{\emph{Acknowledgements.}} The research leading to these results has received funding from the UK Science and Technology Facilities Council (consolidated grants ST/N000609/1 and ST/S000402/1) and the European Union Horizon 2020 research and innovation programme (grant agreement No. 647214). IDM acknowledges support from the Research Council of Norway through its Centres of Excellence scheme, project number 262622. TVD was supported by the
European Research Council (ERC) under the European Union's Horizon 2020
research and innovation programme (grant agreement No 724326) and the C1
grant TRACEspace of Internal Funds KU Leuven

\bibliographystyle{aa}        
\bibliography{ntlw_vrms_paper.bib}           

\end{document}